\documentclass[conference]{IEEEtran}
\IEEEoverridecommandlockouts
\usepackage{cite}
\usepackage{amsmath,amssymb,amsfonts}
\usepackage{algorithmic}
\usepackage{graphicx}
\usepackage{textcomp}
\usepackage{xcolor}
\def\BibTeX{{\rm B\kern-.05em{\sc i\kern-.025em b}\kern-.08em
    T\kern-.1667em\lower.7ex\hbox{E}\kern-.125emX}}

\usepackage{booktabs}
\usepackage[caption=false,font=footnotesize]{subfig}

\begin{document}

\title{Games Mapper: Topological Data Analysis of Steam Genres\\
}

\author{\IEEEauthorblockN{Nicolas Grelier}
\IEEEauthorblockA{Pullup Entertainment \\
Paris, France \\
 nicolas.grelier@pullupent.com}
 \and

 \IEEEauthorblockN{Stéphane Kaufmann}
 \IEEEauthorblockA{Pullup Entertainment \\
 Paris, France \\
 georges.stefanossi@gmail.com}
 
\and \IEEEauthorblockN{Johannes Pfau}
 \IEEEauthorblockA{Utrecht University \\
 Utrecht, Netherlands \\
 j.pfau@uu.nl}
}


\maketitle

\begin{abstract}
The video game industry comprises a vast, continuously evolving landscape of themes and genres. For studios and publishers that navigate this competitive market, understanding the structural dynamics and temporal evolution of specific game categories is crucial for identifying viable entry points. In this paper, we introduce \textit{Games Mapper}, a novel analytical tool based on the Mapper algorithm from topological data analysis. Unlike traditional clustering techniques, Games Mapper captures the continuous topological relationships between datasets over time (or other guiding variables). We extend the standard algorithm with an automated cluster labelling method, ensuring highly interpretable and interactive visualisations of genre evolution. To demonstrate the efficacy of our approach, we present a comprehensive case study on \textit{Simulation} games released on Steam between 2015 and 2025. Games Mapper autonomously segments the genre into coherent, persistent subgenres, and captures dynamic market shifts. 
Ultimately, we provide a scalable, generalisable tool for researchers and industrials to unravel complex market structures and track the evolution of the Steam ecosystem.
\end{abstract}


\section{Introduction}

The video game industry comprises a vast and constantly evolving landscape of genres and subgenres. For studios seeking to enter a new market, understanding the dynamics of these categories (specifically how different subtypes evolve over time, and what barriers to entry might exist) is crucial. Exploring these market structures requires comprehensive data, and platforms like \textit{Steam} provide a rich, well-documented ecosystem complete with user-defined tags, release dates, and engagement metrics. In this paper, we propose a methodology for analysing the structural evolution of any game genre (or subset of games in general), utilising \textit{Simulation Games} as a suitable, comprehensive case study.

Simulation games constitute a popular and highly diverse segment of the industry, ranging from management and city-building to complex physical or exploratory environments. Recent successful releases, such as Supermarket Simulator, Farming Simulator 25, Cities: Skyline 2, and Manor Lords, perfectly illustrate the breadth of this genre. This level of diversity makes it a promising use case to be unravelled by our approach. By examining the different subtypes within the simulation genre and their evolution over time, our goal is to demonstrate the efficacy of our analytical approach while deriving insights into potential barriers to entry for studios seeking to develop their first simulation games.

To conduct this analysis, we base our approach on top of a tool from topological data analysis (TDA), namely the Mapper algorithm~\cite{singh2007topological}. While traditional data analysis methods are commonly used to explore markets and consumer behaviour, the application of TDA for market exploration remains relatively novel. Unlike approaches that primarily rely on geometric representations of data, TDA seeks to reveal the structural and topological relationships within a dataset, making it possible to highlight connections, clusters, and persistent structures that may otherwise remain hidden. The analysis of persistent structures is particularly relevant to the video game industry, where development cycles frequently span three years or more. Consequently, developers and publishers must make strategic decisions based on trends that are likely to persist throughout the development process rather than on transient market fluctuations.

The Mapper algorithm requires the choice of a clustering algorithm as well as several hyperparameters~\cite{singh2007topological}. In this work, we tailor these choices to the study of simulation games; however, the overall methodology can easily be extended to address other questions concerning games released on Steam. In particular, the clustering algorithm and the visualisation procedure can remain unchanged across different genres, while some hyperparameters, such as the filter function and the covering of the filter range, must be adapted depending on the research question.

By deploying the clustering algorithm presented by Grelier and Kaufmann~\cite{grelier2024automated} into the Mapper algorithm, we provide, to the best of our knowledge, a new application of the Mapper algorithm in which the resulting vertices are automatically named according to the dominant tags of the games they contain. This ensures the resulting topological maps remain highly interpretable, regardless of the game genre being analysed. We name this method \textit{Games Mapper}.
To guide our analysis, we formulate the following research questions.

\textbf{RQ1}: Can the genre of simulation games be automatically segmented into coherent persistent subgroups with Games Mapper?

\textbf{RQ2}: How has the market landscape for simulation games (or persistent subgenres) evolved in recent years, particularly regarding release volume and genre diversification?

\begin{figure*}[h!]
    \centering
    \includegraphics[width=\linewidth]{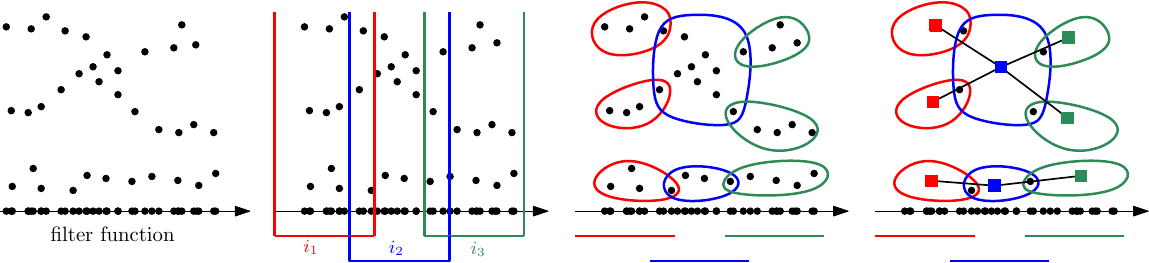}
    \caption{The four main steps of the Mapper algorithm: 1) applying a filter function to project the data, 2) covering the filter range with overlapping intervals, 3) clustering the data within each interval, and 4) connecting clusters that share data points to form the Mapper graph.}
    \label{fig:mapper_def}
\end{figure*}

In \textbf{RQ1}, we focus on identifying subgroups of simulation games that are defined by shared gameplay features. In particular, we are interested in gameplay mechanics and player interactions that characterise different types of simulation experiences. By contrast, we do not consider purely artistic or aesthetic choices (such as graphical style, narrative tone, or visual presentation) unless they correspond to gameplay-related elements. This focus allows us to analyse the structural organisation of simulation games in terms of gameplay design, which is particularly relevant for understanding market segments and potential entry points for new developers.

By incorporating the \textit{Mapper} algorithm from TDA, we aim to identify structural relationships between subgroups of games released in different years and to analyse how these subgroups evolve over time. It explicitly captures overlaps between clusters across consecutive intervals, which makes it possible to track the emergence, persistence, or disappearance of subgenres. More classical data analysis approaches, such as clustering independently for each year or relying solely on dimensionality reduction visualisations, would not naturally capture these structural relationships between groups of games over time.

The contributions of this paper are twofold: First, we provide an empirical analysis of the evolution of simulation games on the Steam platform, identifying the main subcategories of the genre and characterising their temporal dynamics. Unlike previous work, our approach allows the algorithm to discover subcategories autonomously, rather than relying on pre-established classifications, while still enabling a clear visualisation of their evolution over time. Second, we introduce a novel application of the Mapper algorithm in which clusters are automatically assigned interpretable names based on their dominant tags. This methodology can be reused to explore other questions related to the structure and evolution of the Steam ecosystem, beyond the \textit{Simulation} use case -- and even beyond \textit{genres}, which are only \textit{one} way to categorise games. We openly publish the entire approach in the github repository~\footnote{https://github.com/JohannesPfau/GamesMapper} accompanying this paper.

\section{Related Work}
\subsection{Studies of the Steam market}
\textit{Steam} constitutes the largest platform for video game distribution of our time, and has served frequently as a broad, ecologically valid environment of trend and market analysis~\cite{windleharth2016full}.
Approaches to unravel the vast amounts of data are manifold: Most recently, Grelier {\em et al.} present a way of exploring the trends in Steam tags across the years by means of proportional Cohen's h measures (and classify them into ``fads'', ``fashions'' and ``classics'')~\cite{grelier2025analyzing};
Cunha {\em et al.} builds on the median numbers of players of games through time, and cluster the shapes into five groups~\cite{cunha2024shape}; and
Kainz and Pirker introduced a new tool allowing for easy queries and visualisation for predefined analytical jobs~\cite{kainz2023steamvis}. %

As there are no distinct best practices in this field yet, and no published work addressed the benefit of TDA methods for this so far, we are looking forward to contribute insights from our Games Mapper approach.

\subsection{The Mapper algorithm}
The Mapper algorithm is a tool from TDA introduced in 2007 by Singh, M{\'e}moli and Carlsson~\cite{singh2007topological}. It has been recently applied across diverse domains, including physics, biology and finance~\cite{madukpe2026comprehensive}. 
The Mapper algorithm provides a way to visualise high-dimensional data by outputting a graph 
that captures topological features present in a dataset.
Conceptually, the Mapper algorithm proceeds according to the following steps, illustrated in Fig.~\ref{fig:mapper_def}. First, a filter function is applied to the dataset to project high-dimensional data onto a lower-dimensional space (usually one-dimensional). The range of this filter function is then covered with overlapping intervals. Within each interval, a clustering algorithm is applied to the points in that interval, producing a set of local clusters. Finally, clusters from consecutive intervals are connected if they share data points, forming the vertices and edges of the Mapper graph. This construction allows the Mapper graph to reveal topological structures, such as branches, loops, or clusters, that capture the organisation of the data.

Although often referred to as an algorithm, Mapper may be more appropriately described as a framework or methodology, since its application requires multiple design choices beyond merely setting parameter values. For example, the algorithm relies on a clustering procedure, which can range from simple K-means to spectral clustering, DBSCAN, or any other clustering algorithm suitable for the data.
Fitzpatrick {\em et al.} highlight that selecting appropriate hyperparameters and choosing an effective clustering algorithm can be challenging, particularly without comprehensive knowledge of the underlying dataset~\cite{fitzpatrick2025new}. To address this, they propose a method for automatically selecting hyperparameters, making Mapper more accessible to a broader audience of non-specialists in data analysis.

\section{Technical Background}
\subsubsection{Steam Games Clustering}

To join the advancements of TDA and game analytics, we rely on the recent Steam tag clustering method from Grelier and Kaufmann, based on the K-means algorithm~\cite{grelier2024automated}. Their algorithm not only partitions a set of games $S$ into clusters, but additionally names each cluster. For a cluster $C$, which we interpret as the set of games contained in the cluster, and a tag $t$, they denote as $h(t,C)$ the Cohen's h of the proportion of games with the tag $t$ in $C$, denoted as $p_C$, with respect to the proportion of games with the tag $t$ in $S$, denoted as $p_S$. By definition, $h(t,C)$ ranges from $-\pi$ to $\pi$, and a high value means that $p_C$ is significantly larger than $p_S$, and that $p_C$ is rather large in itself~\cite{cohen2013statistical}. In particular, $h(t,C)$ is positive if and only if $p_C$ is larger than $p_S$. The authors suggest to name the cluster $C$ by using the tag $t$ that maximises $h(t,C)$ over all tags. Informally, such $t$ is the most overrepresented tag in the cluster. Secondly, they define the naming score $n(\mathcal{C})$ of a clustering $\mathcal{C}$ with $k$ clusters as a weighted average of the Cohen's h of the best tag in each cluster:

\begin{equation}\label{naming_score}
n(\mathcal{C}) := \frac{1}{|S|} \sum_{1\leq i \leq k} |C_i| \max_t h(t,C_i)
\end{equation}
The naming score also ranges from $-\pi$ to $\pi$, and a high value indicates that each cluster contains an overrepresented tag specific to it, which is desirable. We later apply the elbow method with this naming score of a clustering for selecting the number of clusters in the Mapper algorithm.

\subsubsection{Mapper Graph Visualisation}

Several approaches exist for visualising the Mapper graph. A first question concerns the placement of the vertices. An intuitive solution 
assigns to each vertex the average position of the data points it contains~\cite{dlotko2024mapper}. When the data lie in $\mathbf{R}^2$, the resulting graph can be visualised directly. Otherwise, dimensionality reduction techniques must first be applied, such as PCA~\cite{pearson1901liii}, t-SNE~\cite{hinton2002stochastic}, or UMAP~\cite{mcinnes2018umap}.

Once vertex positions are determined, additional information is often displayed through visual encodings. For example, vertices may be coloured according to a relevant key performance indicator (KPI), or textual information may be displayed inside the vertices themselves~\cite{dlotko2024mapper,fitzpatrick2025new}.

Because edges in the Mapper graph only connect clusters belonging to overlapping intervals of the filter function, the resulting graph naturally has a layered structure. However, Mapper graphs are typically not visualised explicitly as layered graphs, for example by placing all vertices corresponding to the same interval on the same horizontal level. One possible reason is that such a representation may obscure geometric information derived from the original data. 

In the present work, however, clusters are automatically assigned descriptive names. For this reason, readability becomes a primary concern, as the vertex labels must remain easily visible. Moreover, in our setting, the filter function corresponds to the release year of the games, which naturally induces a temporal ordering of the intervals. Representing the Mapper graph as a layered graph therefore facilitates the interpretation of temporal trends. We thus adopt a layered layout in which vertices from the same interval are placed on a horizontal line and spaced evenly. The vertical distance between layers is kept constant and chosen to maximise readability. This approach corresponds to the classical framework of layered graph drawing, which has been explored in~\cite{sugiyama2007methods}.

\section{Case Study: Simulation Games}

To showcase the applicability of our Games Mapper approach, we apply it to simulation games and analyse the resulting topological representation. There is currently no known automatic method for selecting these hyperparameter values, as the state of the art can only help users who are non-specialists in data analysis~\cite{fitzpatrick2025new}. We select ourselves the hyperparameters and the clustering algorithm using our expertise, and provide explanations as to why those choices are relevant. The only hyperparameters we do not select from prior knowledge are the numbers of clusters, in which case we rely on the elbow method.
We mainly follow the four-step procedure of Fitzpatrick {\em et al.}~\cite{fitzpatrick2025new} (\textit{B, C, D, E}) as illustrated in Fig.~\ref{fig:mapper_def},
while outlining two additional stages: an initial step consisting of the construction of the database (\textit{A}), and a final step devoted to the interpretation of the resulting graph (\textit{F}).

\subsection{The database}
We consider games released between 2015 and 2025, with at least $100$ reviews, and that have the tag \emph{Simulation} with a \textit{priority} of at least $0.6$. The \textit{priority} of a tag assigned to a game is a score, ranging from $0$ to $1$, proportional to the number of players that assigned the tag to this specific game. This follows related work on Steam tag analysis~\cite{grelier2024automated}, to exclude hobbyist games or games where simulation is not a primary feature. We consider only sufficiently recent games, as Steam tags were only introduced in 2014~\cite{steam_tags_2014}. 
%
%
The eventual database consists of those selected games (per year) and 
their capital tag priorities, which contain the essential information necessary to describe gameplay features~\cite{grelier2024automated}. 

\subsection{Data filtration}
We define the filter function as the year of release on Steam (official or early access). This choice induces a temporal layering of the Mapper graph and allows us to address \textbf{RQ2}. 
In particular, the resulting structure makes it possible to visualise how subcategories of simulation games evolve over time.
%

\subsection{Data Range Covering}
To construct the covering, we partition the range of the filter function into overlapping intervals. Consecutive intervals are required to overlap, and both the interval length and the overlap size are parameters chosen by the user.
Suppose the filter values lie in the interval $[\min X, \max X]$. We divide this range into $k$ intervals, each of length $\ell$, such that consecutive intervals overlap by length $\varepsilon$, where $0<\varepsilon<\ell$. Since there are $k$ intervals and $k-1$ overlaps, we have $\max X - \min X = k\ell - (k-1)\varepsilon$.
Alternatively, the covering can be defined using the empirical distribution of the filter values rather than their numerical range. In this case, intervals are specified by percentiles. For example, let us fix a parameter $0<\ell<1$ for the interval width in proportion of data, and $0<\varepsilon<\ell$ for the overlap proportion. The first interval contains the lowest $\ell \%$ of the data. The second interval contains the data between $(\ell -\varepsilon) \%$ and $(2\ell -\varepsilon )\%$. In general, each subsequent interval begins $\varepsilon$ percentiles before the end of the previous one.

In our approach, we cover the range of the image of the filter function by setting an interval to be two consecutive years. Two consecutive intervals overlap by one year. We choose this method as it allows for a better interpretability of the results: The time length between layers $i$ and $i+k$ is equal to $k$ years, whatever the value of $i$. Furthermore, if two clusters have the same name, and the most recent one is larger than the other, we can then infer that the number of such games increased. Moreover, the video game industry largely follows yearly cycles, driven by factors such as annual franchises, financial reporting periods, and holiday sales. Defining intervals that coincide with calendar years therefore yields a more natural partition of the data and avoids splitting games that were released under similar market conditions.

As we consider games from $2015$ to $2025$, we have eleven intervals. In the remainder of the paper, we denote as $(i, i+1)$ the layer containing the games released in year $i$ or $i+1$. Note that we do not have games released in 2026 in our database. Thus, the last layer (2025-2026) contains fewer games than the penultimate one (2024-2025). To better remind the reader of this peculiarity of the last layer, we change the notation and denote it as (2025).

\subsection{Clustering}\label{subsec:clustering}
We employ the clustering method proposed in~\cite{grelier2024automated}, as it is specifically designed for Steam game data. This method, based on the K-means algorithm, clusters games according to the similarity of their capital tag priorities and automatically assigns a descriptive name to each cluster. 
In addition, it provides a naming score ranging from $-\pi$ to $\pi$, which quantifies the quality of the clustering. A higher value indicates that clusters are characterised by tags that occur frequently within the cluster but rarely in the others.

The naming score is used to select the clustering resolution using the elbow method, by varying the parameter controlling the number of clusters. As detailed in Subsection~\ref{subsec:visualisation}, we use an heuristic for drawing the Mapper graph in a readable way. To do so, we do some optimisation on each layer, by testing all permutations of the vertices of the layer, where each vertex corresponds to a cluster. Even more, we optimise the first two layers together, by testing all permutations over these two layers. Because of the combinatorial growth of the number of permutations, we impose an upper bound of seven vertices per layer. This ensures that the optimisation procedure remains computationally tractable. Under this constraint, the first two layers may require testing up to $(7!)^2 = 25,401,600$ permutations.

\subsection{Graph construction}\label{subsec:visualisation}
Because each cluster is automatically assigned a name by the clustering algorithm, we display this name above the corresponding vertex: %
for each cluster $C$, the name is the tag $t_{\max}$ that maximises $h(t,C)$~\cite{grelier2024automated}. However, this approach may discard relevant information when several tags have similar $h(t,C)$ values. To address this limitation, we name each cluster using the tags $t$ satisfying $h(t,C) \geq 0.8h(t_{\max},C)$, ordered by decreasing values of $h(t,C)$. For readability, we retain at most the three tags with the highest scores.

The Mapper graph is represented as a layered structure, where vertices corresponding to the same interval are placed on the same horizontal layer. By construction, edges occur only between vertices belonging to consecutive layers.
To improve readability, we position the vertices of each layer so as to minimise the horizontal displacement of the edges. Within a given layer, vertices are constrained to lie on the same horizontal line, evenly spaced and symmetrically arranged about the vertical axis $x=0$. Edges are weighted according to the number of games shared by the two clusters they connect. Since minimising crossings in layered graphs is an NP-hard problem~\cite{eades1994edge}, we adopt a heuristic approach. We first determine the ordering of the vertices in the two lowest layers by testing all permutations and selecting the configuration that minimises the weighted sum, over all edges, of the absolute difference between the horizontal coordinates of their endpoints. We then proceed iteratively, fixing the previous layers and testing all permutations of the next layer above, again selecting the ordering that minimises this quantity. The motivation for this heuristic is that clusters connected by high-weight edges are likely to represent similar groups of games. Minimising the horizontal displacement of edges therefore tends to align related clusters across consecutive layers, making their evolution through time easier to follow.

The reason for jointly optimising the first two layers is that fixing the order of the first layer arbitrarily can lead to layouts that are significantly less readable. In practice, we observed that the resulting drawing was highly sensitive to the ordering of the first layer. Optimising the first two layers simultaneously significantly improves readability and makes the layout deterministic, avoiding poor initial configurations.
As shown in Subsection~\ref{subsec:clustering}, optimising over the first two layers may make us consider $(7!)^2 = 25,401,600$ possibilities, when the first two layers both have seven vertices. We choose to stop there and not optimise over the first three layers together, as it could require to test $(7!)^3 = 128,024,064,000$ possibilities, which was deemed too many.

Vertices are drawn with an area proportional to the number of games that the corresponding cluster contain. In the interactive version of our visualisation (via the \textit{plotly} Python library), hovering over a vertex displays the five games with the highest number of reviews in the cluster. These are typically the titles most likely to be recognised by the reader. Edge darkness and width is also scaled according to edge weight, that is, the number of games shared by the two clusters connected by the edge.

\subsection{Interpretation}
In the resulting Mapper graph, edges are weighted according to the number of games shared by the two clusters they connect. Consequently, edges with higher weights correspond to stronger relationships between clusters. We therefore focus primarily on paths formed by high-weight edges. Long paths of such edges may indicate the persistence of a particular subtype of simulation games over several years. Conversely, when several high-weight edges are incident to a vertex, this may indicate that a subtype branches into multiple related subtypes in subsequent years.

Because the clustering partitions all (simulation) games released within a given time interval into a small number of clusters, these clusters are expected to contain a relatively large number of games and therefore represent relatively broad subcategories rather than highly specialised niches. When clusters with the same name appear across several years, we interpret the corresponding sequence by adopting the terminology introduced in~\cite{bae2025mathematical,grelier2025analyzing}. A subtype is categorised as trending when the number of games increases over time, as a ``fad'' when it corresponds to a short but intense phenomenon, as a ``fashion'' when it persists for two or three years, and as a ``classic'' when it remains present over a long period.

Within each cluster, we examine the games with the most reviews. Since a large number of reviews is generally associated by industry experts with commercial success on Steam~\cite{vginsights2021steam}, this provides an indication of the most successful games within each subtype. We then examine the proportion of indie, AA, and AAA titles among these games in order to address~\textbf{RQ2}. 

\section{Results}

\begin{figure}
    \centering
    \includegraphics[width=\linewidth]{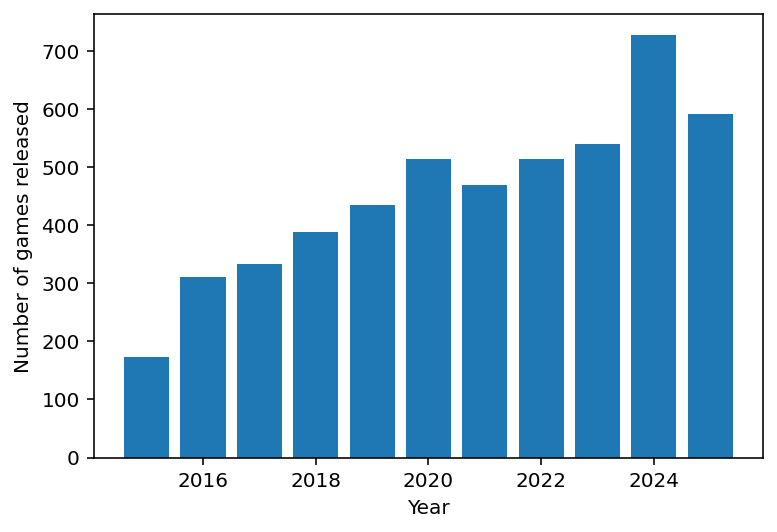}
    \caption{Simulation games released on Steam per year (with $\ge100$ reviews)}
    \label{fig:histogram_years}
\end{figure}

\begin{figure}[!h]
\centering

\subfloat[(2015-2016)]{\includegraphics[width=0.3\textwidth]{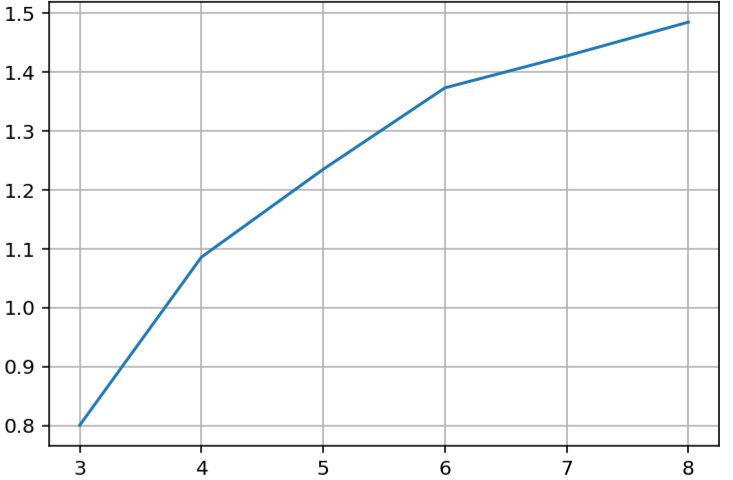}}
\hfill
\subfloat[(2017-2018)]{\includegraphics[width=0.3\textwidth]{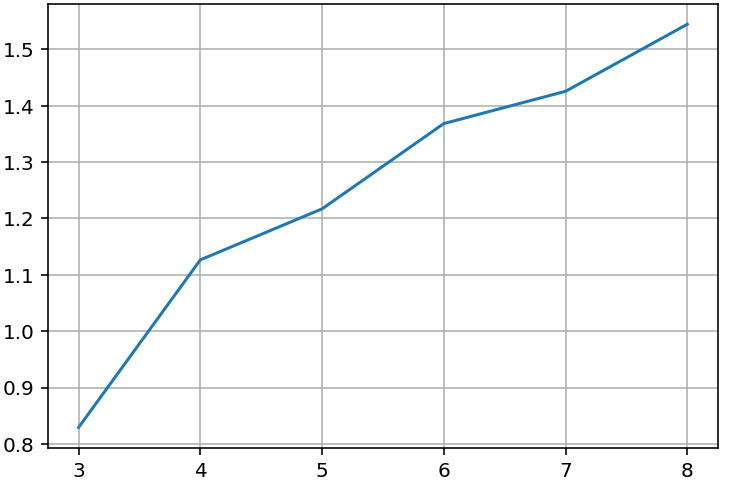}}
\hfill
\subfloat[(2023-2024)]{\includegraphics[width=0.3\textwidth]{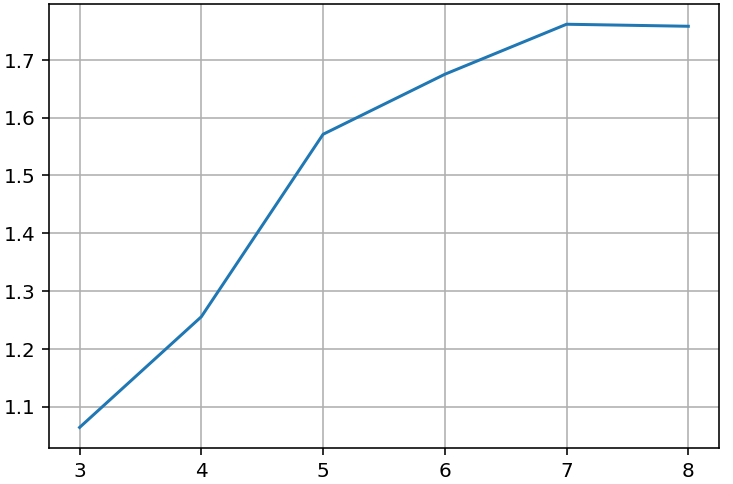}}

\caption{Results for the elbow method: naming scores of the clusterings for some intervals depending on the number of clusters selected.}
\label{fig:elbows}
\end{figure}

\begin{figure*}
    \centering
    \includegraphics[width=1\linewidth, trim = 1cm 2cm 1cm 1cm]{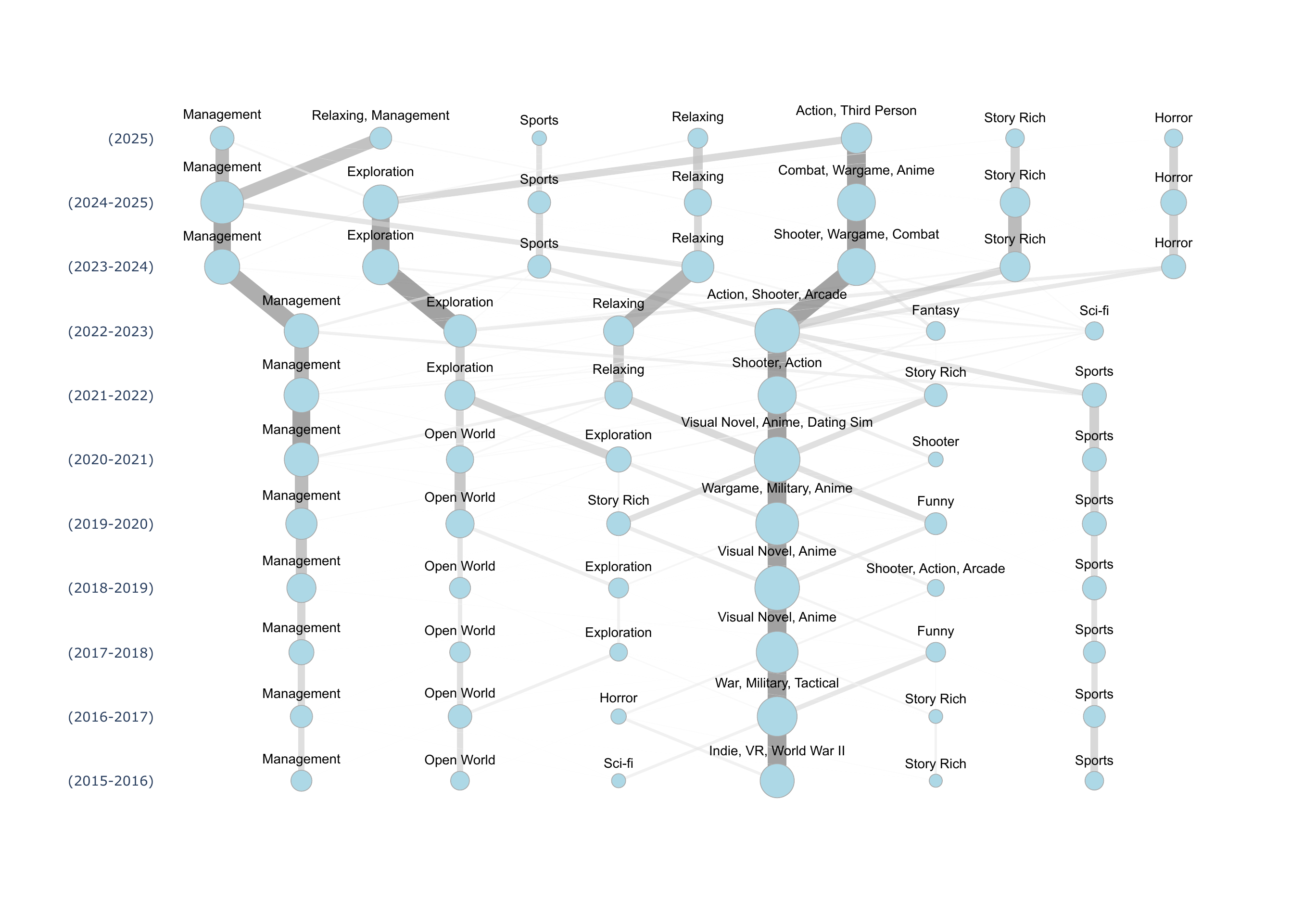}
    \caption{The obtained Mapper graph of simulation games}
    \label{fig:mapper_graph}
\end{figure*}

Fig.~\ref{fig:histogram_years} outlines the distribution of released games per year (as their \textit{filter value}). 
We observe that the numbers are increasing, excepted for the last year. This is probably due to the fact that we only consider games with at least $100$ reviews, and that older games have had more time to get reviews.
We then applied the elbow method to each layer, where a layer contains the games in the preimage of a given interval by the filter function. Fig.~\ref{fig:elbows} demonstrates the naming scores depending on the number of clusters for some intervals. When deciding the number of clusters to be $6$ for the layers ranging from (2015-2016) to (2022-2023) and to be $7$ for the last three layers, the resulting Mapper graph can be visualised as depicted in Fig.~\ref{fig:mapper_graph}. The interactive version of this (which allows contextual information on hover, zooming, highlighting etc.), as well as the entire code base, is available at the open source github repository~\footnote{https://github.com/JohannesPfau/GamesMapper} that accompanies this paper.

\section{Discussion}

\begin{table}
\caption{Tag prevalence within cluster versus the rest of the clusters.}
\label{tab:tag_prevalence}
\centering
\begin{tabular}{llrr}
\toprule
Tag & Interval & In cluster (\%) & Other clusters (\%) \\
\midrule
Management & 2015 & 100.00 & 6.68 \\
Open World & 2015 & 98.44 & 5.48 \\
Sports & 2015 & 98.41 & 1.66 \\
Story Rich & 2015 & 100.00 & 1.55 \\
Sci-fi & 2015 & 100.00 & 4.46 \\
Open World & 2016 & 99.01 & 4.24 \\
Management & 2016 & 100.00 & 8.14 \\
Story Rich & 2016 & 100.00 & 4.11 \\
Horror & 2016 & 97.67 & 2.83 \\
Sports & 2016 & 98.85 & 1.98 \\
Open World & 2017 & 100.00 & 11.32 \\
Funny & 2017 & 100.00 & 7.21 \\
Management & 2017 & 99.12 & 8.88 \\
Exploration & 2017 & 98.25 & 4.36 \\
Sports & 2017 & 98.86 & 2.37 \\
Management & 2018 & 98.70 & 7.17 \\
Sports & 2018 & 99.03 & 1.25 \\
Exploration & 2018 & 98.63 & 5.73 \\
Open World & 2018 & 98.77 & 11.86 \\
Funny & 2019 & 100.00 & 8.83 \\
Management & 2019 & 99.44 & 8.43 \\
Open World & 2019 & 99.31 & 6.72 \\
Story Rich & 2019 & 100.00 & 4.97 \\
Sports & 2019 & 100.00 & 1.19 \\
Open World & 2020 & 100.00 & 7.90 \\
Management & 2020 & 100.00 & 11.67 \\
Sports & 2020 & 99.04 & 1.14 \\
Shooter & 2020 & 100.00 & 1.38 \\
Exploration & 2020 & 100.00 & 9.80 \\
Story Rich & 2021 & 98.94 & 6.97 \\
Management & 2021 & 100.00 & 15.73 \\
Relaxing & 2021 & 100.00 & 13.36 \\
Sports & 2021 & 99.05 & 1.25 \\
Exploration & 2021 & 83.44 & 6.59 \\
Sci-fi & 2022 & 100.00 & 2.11 \\
Fantasy & 2022 & 100.00 & 4.04 \\
Exploration & 2022 & 100.00 & 5.91 \\
Management & 2022 & 100.00 & 17.04 \\
Relaxing & 2022 & 100.00 & 12.81 \\
Exploration & 2023 & 99.58 & 11.25 \\
Management & 2023 & 98.66 & 19.46 \\
Horror & 2023 & 99.06 & 3.53 \\
Relaxing & 2023 & 100.00 & 15.43 \\
Sports & 2023 & 100.00 & 1.20 \\
Story Rich & 2023 & 100.00 & 10.42 \\
Management & 2024 & 98.79 & 13.77 \\
Relaxing & 2024 & 100.00 & 21.86 \\
Sports & 2024 & 100.00 & 1.79 \\
Horror & 2024 & 99.17 & 3.76 \\
Exploration & 2024 & 99.55 & 8.56 \\
Story Rich & 2024 & 100.00 & 11.82 \\
Story Rich & 2025 & 100.00 & 11.53 \\
Relaxing & 2025 & 100.00 & 22.12 \\
Management & 2025 & 99.03 & 24.54 \\
Sports & 2025 & 100.00 & 2.17 \\
Horror & 2025 & 98.33 & 3.01 \\
\bottomrule
\end{tabular}
\end{table}

To address \textbf{RQ1}, our Games Mapper successfully segmented the simulation genre into coherent subgroups, as evidenced by the persistence of clusters such as \textit{Management} and \textit{Sports}. Table~\ref{tab:tag_prevalence} reports, for each cluster named with a single tag $T$ over the interval $(Y, Y+1)$, two quantities: the percentage of games \emph{within} that cluster that have the tag $T$, and the percentage of games in the \emph{remaining} clusters of the same interval that have $T$. For readability, we label each interval by $Y$ rather than by $(Y, Y+1)$. We restrict this analysis to clusters named with a single tag, for which the results are more directly interpretable than for clusters named with several tags.

Nearly all games in each single-tag cluster have the corresponding tag, and in the few clusters where this does not hold exactly, the percentage remains very close to $100\%$. This indicates a high intra-cluster similarity. Conversely, the share of games in the other clusters that have a given cluster's tag is low, often below $10\%$. We do not regard the fact that this share is not extremely close to $0\%$ as problematic: A game may legitimately have tags associated with several clusters while being assigned to only one, so it is expected that games outside a cluster occasionally share its tag. Consistently with this, the clusters with the highest such percentages are \emph{Open World}, \emph{Exploration}, \emph{Management}, and \emph{Relaxing}, i.e. tags that are applied broadly. Many horror or shooter games, for instance, may naturally also be tagged \emph{Open World} or \emph{Exploration}. Taken together, these percentages remain low, indicating a low inter-cluster similarity. We conclude that the clustering component of the Mapper algorithm performs as intended.

We now exhibit the insights provided by the topology visualised in the Mapper graph. Without topology, one could still partition the games into yearly intervals, and independently cluster the games within each interval. Such an approach would identify groups of similar games at each point in time, but it would not provide explicit information about the relationships between clusters across consecutive intervals. In contrast, the edges of the Mapper graph encode these relationships and facilitate the interpretation of temporal dynamics.

For example, when a cluster with a given name appears in every interval, as is the case for \textit{Management} in Fig.~\ref{fig:mapper_graph}, one can already infer that the corresponding subgenre is persistent. However, Games Mapper reveals additional information through the edge structure. The edges connecting consecutive \textit{Management} clusters have substantially larger weights than the other edges incident to these vertices, indicating not only persistence but also a strong continuity through time. Thus, management games form a relatively isolated and stable subgroup within the broader simulation genre.

Conversely, the topology also highlights clusters that do not emerge from a clearly identifiable lineage. Consider the \textit{Sci-fi} cluster in the interval (2022-2023). Although it is connected to several clusters in the adjacent intervals, none of these connections has a particularly large weight. This suggests that the cluster neither originates from a dominant preceding trend nor evolves into a dominant subsequent trend.

Finally, topology helps identify relationships between distinct subgenres. In three intervals, both an \textit{Open World} cluster and an \textit{Exploration} cluster are present. By examining the strongest edges incident to the \textit{Exploration} clusters, we observe that they consistently connect to \textit{Open World} clusters in adjacent intervals. This indicates a strong proximity between these two subgenres.

Regarding \textbf{RQ2}, we interpret the outcomes of the final visualisation outlined in Fig.~\ref{fig:mapper_graph} regarding multiple evolving cluster names.
We observe that the \emph{Management} category appears in every interval, while \emph{Sports} is present in all but the (2022–2023) interval. Beyond these, we interpret the Mapper graph by following sequences of vertices connected by high-weight (darker with larger width) edges. In the remainder of this paper, we refer to these sequences of clusters connected by high-weight edges as ``cluster trajectories''. As discussed in Subsection~\ref{subsec:visualisation}, we introduced a heuristic designed to improve the readability of the Mapper graph by reducing the horizontal distance between connected clusters. We observe that the resulting cluster trajectories are indeed largely vertical, which makes their evolution through time easy to follow.

We notice a cluster trajectory that transitions from \emph{Open World} to \emph{Exploration}. In addition, there are shorter cluster trajectories that do not span the entire 2015–2025 period, such as \emph{Horror}, \emph{Story Rich}, and \emph{Relaxing}.

\subsection{Management games}
The \emph{Management} category is present in all intervals and shows a clear increase in size over time, going from $80$ games in the interval (2015-2016) to to $331$ in (2024-2025), and contributing significantly to the overall growth in the number of simulation games. The clusters named \emph{Management} include titles like Cities: Skyline I and II, RimWorld, Crusader Kings III, Manor Lords and Schedule I. Consistently, these games have \emph{Management} among their most prominent tags, confirming the validity of the cluster labelling. As can be interpreted from the Games Mapper graph, \textit{Management} simulation games are currently trending. We note that, interestingly, a lot of the most reviewed games are published by AA companies, whereas indie and AAA games are less represented.

\subsection{Open world and exploration games}
The \emph{Open World} category also shows a clear increase in size over time and eventually transitions into \emph{Exploration}, going from $64$ games in (2015-2016) to 221 in (2024-2025). The strong connections between these clusters suggest that, among simulation games, the relevance of the \emph{Open World} tag is gradually replaced by \emph{Exploration} around 2021. Notably, both tags co-exist in separate clusters during the intervals (2017–2018), (2018–2019), and (2020–2021). This observation raises the question of whether the underlying feature remains the same but players have changed how they describe it, or whether the feature itself has evolved.

As per Adams' \textit{Fundamentals of Game Design}, an \textit{Open World} game ``allows players to roam freely through a continuous virtual environment''~\cite{adams2014fundamentals}. Yet, the design of an open world is rather a \textit{dynamic} (after Hunicke {\em et al.}'s MDA framework \cite{hunicke2004mda}), whereas \textit{Discovery/Exploration} is rather an \textit{aesthetic} - a core player motivation and emotional response. Thus, technically, a game could have one without the other.
%
%
To shed more light on this, we manually consider the games in the clusters named \emph{Exploration}, and check if there are games where exploration is a feature while not being an open world. In these clusters, we find: Disney Dreamlight Valley, Dinkum, Strange Horticulture, Abiotic Factor, Pacific Drive. Those games have the tag \emph{Exploration} but not the tag \emph{Open World}. We also find games that have both, like: Microsoft Flight Simulator 2024, Kerbal Space Program 2, SnowRunner and Police Simulator: Patrol Officers. We even find games that don't have the tag \emph{Exploration}, but that were put in the cluster due to their similarity with exploration games. Those that we eventually considered do have the tag \emph{Open World} (Gray Zone Warfare, Ranch Simulator: Build, Hunt, Farm).

Overall, this analysis confirms a strong similarity between open world and exploration games. These games are currently trending. The most reviewed titles in these clusters are typically produced by larger indie studios or AA developers and publishers. It is unsurprising that smaller indie studios are underrepresented, given the substantial development resources required to produce open world games. However, the scarcity of AAA titles in this category is notable, as one might have expected more representation from major studios.

\subsection{Sport games}
We observe that in each interval, there is a cluster named \emph{Sports}, excepted for the interval (2022-2023). The size of these clusters remains relatively constant over time after a first bump from $63$ in (2015-2016) to $87$ games in (2016-2017), and finally $93$ games in (2024-2025), suggesting that the number of sports simulation games has not experienced substantial growth. We find this particularly interesting when put in contrast to management or open world/exploration games, whose numbers kept increasing - whereas sport games seem to be a ``classic'' trend~\cite{grelier2025analyzing}.
Influential sport games in these clusters include the DiRT, F1, NBA, FIFA, Asseto Corsa and Football Manager series. The absence of a sports cluster in the (2022–2023) interval likely reflects the natural randomness of game release schedules and the effects of our clustering algorithm rather than any systematic trend. 
It is notable that the most reviewed titles in these clusters come from major mainstream franchises developed and published by AAA studios, suggesting that the barrier to entry for smaller studios in the sports simulation genre is particularly high.

\subsection{Horror games}
Since 2023, we observe a cluster trajectory of horror games. There was also a cluster named \emph{Horror} in the interval (2016-2017), which was relatively small compared to the other clusters from that interval. In that cluster, we find games like: Rising Storm 2: Vietnam, Freddy Fazbear's Pizzeria Simulator, Job Simulator, SIMULACRA and Cockroach Simulator. Those games all have the tag \emph{Horror}. Some have horror as a main feature, like SIMULACRA, and some have it as a secondary feature, like Rising Storm 2: Vietnam. The case of Job Simulator might be debatable, as this is rather likely tagged from the user community without serious intent.

To know if the existence of an \emph{Horror} cluster in that interval is really significant (implying that there was a fad in simulation horror games), or if it was just a temporary event due to the randomness of release dates and of the clustering algorithm, we consider the clusters from the intervals above and below that are strongly connected to it: \emph{Indie, VR, World War II} in the interval (2015-2016) and \emph{Visual Novel, Anime} in the interval (2017-2018).
In the first interval, we find Viscera Cleanup Detail (horror). In the more recent interval, we find again Rising Storm 2: Vietnam and Freddy Fazbear's Pizzeria Simulator. Thus, it seems that there are truly more horror games in the interval (2016-2017), but those games do not have a very large number of reviews, and do not seem to have had a great impact on the genre. 
In the recent \emph{Horror} clusters, the list of games include: Slay the Princess — The Pristine Cut, FIVE NIGHTS AT FREDDY'S: HELP WANTED, CloverPit, The Children of Clay and That's not my Neighbor. Hence, the genre of simulation horror games appear to be currently trending since 2023, from $106$ games in (2023-2024) to $121$ games in (2024-2025). We also note that most of the highly reviewed games are from indie studios.

\subsection{Story Rich games}
We observe \emph{Story Rich} clusters being present in the intervals (2015-2016), (2016-2017), (2021-2022), (2023-2024), (2024-2025) and (2025). In the first two intervals, the \emph{Story Rich} clusters are the smallest ones, indicating that it was not a strong phenomenon. In those clusters, we find games like VA-11 Hall-A: Cyberpunk Bartender Action, Her Story, Beholder and Emily is Away Too. Apart from the first one, these games have fewer than $5,000$ reviews, confirming that those games did not correspond to mainstream events.
In the interval (2021-2022), we find: Tale of Immortal, Touhou Mystia's Izakaya and Strange Horticulture. The cluster is again the smallest in the interval. In the last three clusters, we have games like: The Operator, FINAL FANTASY TACTICS - The Ivalice Chronicles, Wanderstop and The Hundred Line -Last Defense Academy-. One could argue that story rich simulation games are currently trending, but the absence of such a cluster in the interval (2022-2023) make it hard to conclude with confidence. We acknowledge that most Story Rich games usually occur in other meta-genres (apart from Simulation), and observe that most highly reviewed games are made by indie studios.

\subsection{Relaxing games}
Since 2021, we observe a cluster trajectory of \emph{Relaxing} games. Those games include: DAVE THE DIVER, PowerWash Simulator 1 and 2, Potion Craft: Alchemist Simulator, Coral Island, Farming Simulator 25, inZOI and Cast n Chill. The size of the clusters is staying constant, from $137$ games in (2021-2022) to $134$ games in (2024-2025), indicating that relaxing simulation games form a trend that is apparently becoming a ``classic''. We observe the presence of many indie (and a few AA) games among the ones with the most reviews.

\subsection{War, visual novel and anime games}
The last remaining cluster trajectory has very large clusters. When using the K-means algorithm, it is quite common to have one cluster that is not as homogenous as the others~\cite{vijay2017variance}. This cluster trajectory is the hardest one to interpret. In the earliest intervals, the names oscillate between tags referring to war, e.g. \emph{War}, \emph{Wargame}, \emph{Shooter}, \emph{Military}, and tags referring  to visual novels or anime, e.g. \emph{Visual Novel}, \emph{Anime}, \emph{Dating Sim}. However, since the interval (2021-2024), the names are constantly referring to war.

The games related to war include: World of Warships, Stellaris, Totally Accurate Battle Simulator, Hell Let Loose, Arma Reforger, Mechabellum, Europa Universalis V and Sid Meier's Civilization VII. We observe that most of these highly reviewed games are AAA.
The visual novel or anime games include: Chinese Parents, Touhou Mystia's Izakaya and Slay the Princess. Although several clusters bear those tags as name, the most reviewed games in these clusters are actually containing a majority of war games. Those clusters must still contain mostly visual novel or anime games, but those have fewer reviews than the war games present in the clusters.

A natural way to further investigate this cluster trajectory would be to apply an additional clustering step in order to obtain a finer subdivision. We leave this analysis for future work. The size of these clusters remains relatively stable over the years, going from $287$ games in (2016-2017) to $259$ games in (2024-2025). However, since they contain a heterogeneous mix of games, it is difficult to interpret them clearly or to determine whether they correspond to a stable subgenre in the sense of a ``classic''. Nevertheless, we observe that the visual novel titles appearing in these clusters are predominantly developed by indie studios, whereas the war-themed games are mostly produced by AA or AAA developers.

\subsection{Miscellaneous}
There are a few clusters which we did not mention yet, because they appeared less often than the others. For instance, there are two clusters \emph{Sci-fi}: one in (2015-2016), the other in (2022-2023), which include: Stellaris, Anno 2205, Dune: Spice Wars and I Was a Teenage Exocolonist.
Two \emph{Funny} clusters appear as well, one in (2017-2018) and the other in (2019-2020). They contain: Hand Simulator, Chinese Parents, Loading Screen Simulator and Totally Accurate Battle Simulator.
Eventually, there is a single cluster \emph{Fantasy} in (2022-2023), that includes: Potionomics and Bronzebeard's Tavern.

For these isolated clusters, the available data does not support strong conclusions. In particular, the sample size is too limited to reliably determine which categories of games (indie, AA, or AAA) tend to accumulate a large number of reviews. Although these clusters could reflect temporary trends or fads, the evidence remains insufficient to substantiate such an interpretation.

\section{Limitations and future work}
To focus on games that were developed for commercial purposes rather than by hobbyists, we restricted our dataset to titles with at least $100$ Steam reviews. However, this threshold is arbitrary and introduces a potential bias: Older games are more likely to have accumulated more than $100$ reviews than recently released ones. A possible alternative would be to consider only games that reached $100$ reviews within a fixed time window after release, for example one year. Unfortunately, we do not have access to such data.

In addition, our clustering pipeline relies on the elbow method combined with the K-means algorithm. Both components introduce a degree of stochasticity into the analysis, which may lead to small variations in the resulting clusters. Consequently, some observations, such as the absence of a cluster labelled \emph{Sports} in the interval (2022–2023), cannot be interpreted as definitive evidence of a structural change in the dataset, as they may partly result from the variability of the clustering process.
Another limitation of our analysis lies in the reliance on Steam tags, which are partly user-generated and may therefore reflect community perception rather than the intended design of the games.

In this study we used the release year as the filter function for the Mapper algorithm in order to investigate the temporal evolution of simulation games. Future work could explore alternative filter functions, such as the number of reviews, the proportion of positive reviews, the maximum number of concurrent users, the release price, or the priority of specific tags. Such analyses could provide complementary insights into the structure and dynamics of the Steam ecosystem.

\section{Conclusion}
We presented \textit{Games Mapper}, a new application of the Mapper algorithm extended by an automatic naming technique for its clusters. This algorithm can be applied to any set of Steam games, and to answer many questions, by adapting the filter function and data range covering. To showcase the efficacy of this method, we demonstrated its validity by applying it to simulation games as a case study.

Our analysis shows that simulation games are currently experiencing significant growth, as reflected by the increasing number of such titles released each year. Within this broader genre, subgenres associated with \emph{Management}, \emph{Exploration}, and \emph{Horror} appear to be growing particularly rapidly. In contrast, \emph{Sports} and \emph{Relaxing} games exhibit a more stable presence over time and could be regarded as \textit{classics}. We also observe a gradual transition from \emph{Open World} games to \emph{Exploration} games, suggesting a shift in how players and developers describe similar gameplay features.

Our results also highlight differences in the distribution of studios across subgenres. It appears that the most reviewed \textit{sports} and \textit{war} games are in the preserve of AAA studios and publishers, suggesting that these markets present significant barriers to entry for indie and AA developers. In contrast, among highly reviewed simulation games, AA studios appear most prominently in the \emph{Management} and \emph{Exploration} subgenres. Interestingly, we observe relatively few AAA titles among successful exploration games, which may indicate that this segment is particularly favourable to mid-sized studios.

Finally, indie developers dominate several smaller subgenres, including \emph{Horror}, \emph{Story Rich}, \emph{Relaxing}, \emph{Visual Novel}, and \emph{Anime}. This suggests that these niches may be less attractive for large AAA productions, potentially because their audiences are smaller or more specialised. Overall, these observations provide insights into how different types of studios position themselves within the simulation game market.

\bibliographystyle{IEEEtran}
\bibliography{bib}

\end{document}